\documentclass[pra,twocolumn,floatfix]{revtex4}
\usepackage{amsmath,amssymb,amsthm}
\usepackage{graphicx}
\usepackage[caption=false]{subfig}
\usepackage{bbm}

\begin{document}
\title{Comment on \lq\lq{}Optimal probe wave function of weak-value amplification\rq\rq{}}
\author{Antonio \surname{Di Lorenzo}}
\affiliation{Instituto de F\'{\i}sica, Universidade Federal de Uberl\^{a}ndia,\\
 38400-902 Uberl\^{a}ndia, Minas Gerais, Brazil}
\begin{abstract}
First, a misconception about the spectrum of a confined particle is evidentiated.  
Then, the results are shown to be incorrect by means of a counter-example, 
an explicit preparation for the probe is given that yields an arbitrary amplification, 
and the source of the error is pointed out. 
\end{abstract}
\maketitle
The authors of Ref.~\cite{paper} claim to have determined the optimal probe wave-function that provides the  
maximum average readout for fixed preparation and post-selection. 
The results are independent of the actual coupling strength and they apply to operators satisfying $\Hat{A}^2=1$. 
For clarity and brevity we shall prepend the acronym SSH when referring to equations in Ref.~\cite{paper}.

After Eq. (SSH18), which we should take momentarily as good, Ref.~\cite{paper} states that 
\lq\lq{}the position $q$ of the probe takes the discrete value
$q = 2gn$, with $n$ being an integer because of the boundary
condition in the momentum space.\rq\rq{}
This misleading statement is repeated in the caption of Fig. 2, where the function $\tilde{\xi}_i(q)$ 
is plotted at discrete points. 
This is the complementary form of the common misconception that the momentum of a confined particle is discrete. 
One should not confuse the indices of the Fourier basis for the expansion in energy eigenstates with the allowed values 
of the variable $q$ or $p$. The position $q$ has a continuous spectrum, even if the momentum 
is bounded to the interval $[-\pi/2g,\pi/2g]$.
To make this point more clear, let us consider a particle of mass $m$ confined in a one dimensional box with a coordinate $q\in[0,L]$. 
Thus, the roles of $q$ and $p$ are temporarily switched in order to provide a familiar case. 
It is well known that the energy is quantized according to $E_n = p_n^2/2m$, 
with $p_n =\hbar \pi n/L$. By analogy with the classical expression, it is tempting to identify $p_n$ with the allowed values of the particle\rq{}s momentum. 
However, a von Neumann measurement of the operator $\Hat{P}$ does not necessarily yield a value $p_n$, but has a finite probability for any $p$. 
Even if the wavefunction is $\psi_n\propto\sin{(p_n q)}$, its Fourier transform $\tilde{\psi}_n(p)$ 
gives a continuous probability distribution $|\tilde{\psi}_n(p)|^2$ that has two main peaks at $p_n$ and $-p_n$ but also several secondary peaks. 

Going back to the case studied in Ref.~\cite{paper}, 
it is more convenient to work with the rescaled quantities $x=q/g$ and $k=gp$, so that the pointer shift is 
$\langle \Hat{x}\rangle_f =\langle \Hat{q}\rangle_f/g =(1+|A_w|^2)/2\,\mathrm{Re}A_w$. 
Then the Fourier transform of 
Eq.~(SSH15) yields $\tilde{\xi}_i(x)=\tilde{\xi}_0(x-\langle \Hat{x}\rangle_f)$, 
 where
\begin{align}
\tilde{\xi}_0(x)
=&
\frac{\sqrt{2|\mathrm{Re}{A_w}|}}{\pi (1+\sigma A_w)}
\cos\left(\frac{\pi x}{2}\right)\! \Phi\!\left(\!z,1,\frac{1+\sigma x}{2}\!\right),
\end{align}
with $\sigma$ the sign of $\mathrm{Re}(A_w)$, $z=(1-\sigma A_w)/(1+\sigma A_w)$, and 
$\Phi(z,s,x)=\sum_{k=0} z^k/(x+k)^s$ the Lerch transcendent. 
In particular, at $x=2n$, 
\begin{align}
\nonumber 
\tilde{\xi}_0(2n)=&\frac{ \sqrt{8|\mathrm{Re}{A_w}|}}{\pi (-z)^n(1+\sigma A_w)}
\biggl[\frac{\mathrm{Atanh}{(\sqrt{z})}}{\sqrt{z}}
\\
&\quad\quad\quad -\mathrm{sgn}{(n)}\!\!\!\!\!\sum_{\min[0,n]}^{\max[n-1,-1]}  \frac{z^k}{2k+1}
\biggr]
\end{align}
for any integer $n$, 
while $\tilde{\xi}_0\left(-\sigma(2n+1)\right)=(-z)^n \sqrt{2|\mathrm{Re}{A_w}|}/(1+\sigma A_w)$, 
and $\tilde{\xi}_0\left(\sigma(2n+1)\right)=0$ for $n$ a positive integer.
A further confirmation that $\tilde{\xi}_i$ should not be taken at discrete values is 
that $\sum_n |\tilde{\xi}_i(2n)|^2=1/2$, while $\int dx |\tilde{\xi}_i(x)|^2=1$, so that $|\tilde{\xi}_i(2n)|^2$ do not represent probabilities. 

On the other hand, the claim that $\Delta\langle\Hat{x}\rangle=(1+|A_w|^2)/2\,\mathrm{Re}A_w$ is the maximum pointer shift is unfounded. 
As a counter-example, let us consider a weak measurement of a spin 1/2 along the $X$-direction (no relation to the $x$ representing the pointer readout), 
where the spin is initially prepared along the positive $Z$ axis and afterward post-selected in a direction along the $XZ$ plane  
forming an angle $\theta$ with $Z$. For an ideally weak measurement, obtained in the limit $W\to 0$  of Eq. (SSH14), 
we have $\Delta\langle\Hat{x}\rangle_W=\mathrm{Re} A_w = \sin{(\theta)}/(1+\cos{\theta})$. For an ideally strong 
measurement, instead, ($W\to\infty$) $\Delta\langle\Hat{x}\rangle_S=2\mathrm{Re} A_w/(1+|A_w|^2) = \sin{(\theta)}$. 
For the allegedly optimal preparation, $\Delta\langle\Hat{x}\rangle_O=1/\Delta\langle\Hat{x}\rangle_S = 1/\sin{(\theta)}$. 
In the range $\pi/2\le \theta \le \pi$, the inequality $\Delta\langle\Hat{x}\rangle_S\le \Delta\langle\Hat{x}\rangle_O\le \Delta\langle\Hat{x}\rangle_W$ holds, showing that the probe preparation is not optimal.

We now proceed to show how to obtain an arbitrary shift by properly preparing the probe. 
We consider the functions 
\begin{equation}
\label{guess}
\xi_i(k) = \frac{e^{-i\alpha G(k)}}{B(k)} ,
\end{equation}
with support in $[-n\pi/2,n\pi/2]$, $n\in\mathbb{N}$, 
where $\alpha$ is an arbitrary real number and $G(k)$ is a primitive of $|B(k)|^{-2}$, i.e. $G'(k)=|B(k)|^{-2}$. 
For $n\to \infty$ the function is not normalizable, but gives rise to a finite shift.  
The initial and final shift of the probe are
\begin{equation}
\langle\Hat{x}\rangle_i = (\alpha-\mathrm{Re}A_w) \frac{1+|A_w|^2}{2(\mathrm{Re}A_w)^2} ,
\end{equation}
and 
\begin{equation}
\langle\Hat{x}\rangle_f = -\alpha\frac{1}{|\mathrm{Re}A_w|} ,
\end{equation}
so that 
\begin{equation}
\Delta \langle\Hat{x}\rangle = \frac{1+|A_w|^2}{2(\mathrm{Re}A_w)}+\alpha 
\frac{(1-|\mathrm{Re}A_w|)^2+(\mathrm{Im}A_w)^2}{2(\mathrm{Re}A_w)^2} .
\end{equation}
We notice that the coefficient of $\alpha$ is zero only in the trivial case $A_w=\pm 1$ (when 
the shift is $\Delta\langle\Hat{x}\rangle=\pm 1$ for any initial preparation of the probe, see the analysis below for a discussion 
of this case), otherwise it is always strictly positive. Hence, by properly choosing $\alpha$, $\Delta \langle\Hat{x}\rangle$
can be made equal to any value. 

A careful analysis reveals the following errors in Ref.~\cite{paper}:
\begin{enumerate}
\item 
As Ref.~\cite{paper} is implicitly looking for an extremal of the action (which is called a Lagrangian therein) for fixed values of $\xi_i(k)$ at $k_-,k_+$, and these extrema of 
integration may be finite, 
Eq. (SSH23) and the first line of Eq. (SSH29) are not in general the correct definitions of the averages. They should be replaced by the real expression
\begin{align}
\nonumber
\langle\Hat{x}\rangle =&-\mathrm{Im}{\int dk\, \xi(k)^*  \xi(k)\rq{}}\\
=&i{\int dk\, \xi(k)^*  \xi(k)\rq{}}-\frac{i}{2} \left[|\xi(k)|^2\right]_{k=k_-}^{k=k_+},
\end{align}
with $\xi=B \xi_i$ and $\xi=\xi_i$, respectively.
Consequently, the reasoning that fixes the interval of integration to $[-\pi/2,\pi/2]$ does not hold. 
However, since $|\xi_i|^2$ is periodic with period $\pi$, this error has no consequence for the result.  
\item The integrand contains $\xi_i'$ which has a divergence at $k=\pm \pi/2$ for the wave-function given in Eq. (SSH15).
As the $\delta$-function given by $\xi_i'$ contributes only a real term to $\xi(k)^*  \xi(k)\rq{}$, this error 
has no consequences as well. 
\item Reference \cite{paper} finds an extremum for $\langle\Hat{x}\rangle_f$, not for $\langle\Hat{x}\rangle_f-\langle\Hat{x}\rangle_i$. 
True, one can shift the initial wavefunction in order to have $\langle\Hat{x}\rangle_i=0$, however the variation 
is being made over all wave-functions. As shown below, this error is the cause of the incorrect claim.
\end{enumerate}

\appendix
\section{Correct treatment of the variational problem}
The wave-function of the probe after the measurement is 
\begin{equation}
\langle k|f\rangle=\xi_f(k)/N_f^{1/2} =B(k) \xi_i(k)/N_f^{1/2}
\end{equation}
with the normalization
\begin{equation}
N_f=\int |B(k) \xi_i(k)|^2 dk .
\end{equation}
We shall consider that $\xi_i(k)$ may not be normalized, so that the initial normalized wave-function is $\langle k|i\rangle=\xi_i(k)/N_i^{1/2}$
with the normalization
\begin{equation}
N_i=\int |\xi_i(k)|^2 dk .
\end{equation}
The quantity to be maximized is the pointer shift (from now on we omit the functional dependence on $k$)
\begin{align}
\nonumber
\Delta\langle\Hat{x}\rangle =& \langle f|\Hat{x}|f\rangle - \langle i|\Hat{x}|i\rangle \\
\label{act}
=& -\mathrm{Im}\left\{\frac{\int dk \left[B\xi_i \right]^*  \left[B\xi \right]\rq{}}{N_f}-\frac{\int dk\, \xi_i^*  \xi_i\rq{}}{N_i}\right\}.
\end{align}
The problem is well posed when either the integration is over a finite interval $k_-,k_+$ with $\xi_i$ fixed at the extrema, or when the integration is over the real axis and 
it is required that $\lim_{k\to\pm\infty}\xi_i=0$, with the vanishing of $\xi_i$ fast enough to guarantee that $N_i$ is finite. 


Since the premises of Ref.~\cite{paper} appear problematic, let us rederive the extremal conditions for $\Delta \langle\Hat{x}\rangle$. 
We start by noticing that the action $\Delta \langle\Hat{x}\rangle$ is invariant under the operation $\xi_i(k)\to C e^{ix_0k} \xi_i(k)$ for arbitrary $C\in \mathbb{C}, x_0\in\mathbb{R}$. 
Thus the solution to the extremal problem, if it exists, is not unique. Physically, this symmetry means simply that shifting the initial wave-function $\tilde{\xi}_i(x)$ by $x_0$ 
does not change the shift of the pointer. 
By varying the action, we get 
\begin{align}
\nonumber
&0=\frac{\delta \Delta \langle\Hat{x}\rangle}{i\delta \xi_i^*} \\
&=\frac{B^* [B\xi_i]\rq{}}{N_f}-\frac{\xi_i\rq{}}{N_i}+i\left[\frac{\langle\Hat{x}\rangle_f}{N_f} |B|^2
-\frac{\langle\Hat{x}\rangle_i}{N_i} \right]\xi_i.
\end{align}
Hence 
\begin{align}
\label{eq:diffeq}
&\left[|B|^2-\overline{|B|^2}\right]\xi_i\rq{}
=-\left[B^* B\rq{}+i\langle\Hat{x}\rangle_f|B|^2-i\langle\Hat{x}\rangle_i \overline{|B|^2}\right] \xi_i,
\end{align}
with $\overline{|B|^2}=N_f/N_i$. 
We notice that Eq.~\eqref{eq:diffeq} differs from Eq.~(SSH25), since the extra term originating from $-\langle\Hat{x}\rangle_i$ in the action is missing in Ref.~\cite{paper}.

A special case is $A_w=\pm 1$, which gives $B=\exp{[\mp i k]}$. Then the coefficient of $\xi_i\rq{}$ in Eq.~\eqref{eq:diffeq} vanishes identically, and the acceptable solution is 
$\Delta \langle\Hat{x}\rangle=\pm 1$, with $\xi_i$ arbitrary. This has a simple physical interpretation: if $B(k)=\exp{[\mp i k]}$, the effect of the interaction (which, we remind the reader, is of multiplying the initial wave-function by $B$)  in the position representation is to shift the wave-function by $\pm 1$, so that for any initial state, the shift is always the same. 

For $A_w\neq\pm 1$, we have a first order equation, with singular points (the continuous function $|B|^2$ certainly equals its average for some $k$). 
The quantities appearing in the equation are
\begin{align}
\nonumber
|B|^2=& \frac{1}{2}\left(1+|A_w|^2\right)+\frac{1}{2}\left(1-|A_w|^2\right)\cos{(2k)}\\
&+\mathrm{Im}(A_w) \sin{(2k)}
\end{align}
and 
\begin{align}
\nonumber
B^* B' =& -\frac{1}{2}\left(1-|A_w|^2\right)\sin{(2k)} + \mathrm{Im}(A_w) \cos{(2k)}\\
&-i\mathrm{Re}(A_w) .
\end{align}
We remark that $2\mathrm{Re}(B^* B')= \tfrac{d |B(k)|^2}{dk}$
The differential equation Eq.~\eqref{eq:diffeq} is hence rewritten as 
\begin{align}
\frac{\xi_i'}{\xi_i} =& -\frac{D'}{2 D}-i\langle\Hat{x}\rangle_f-i \frac{\Delta\langle\Hat{x}\rangle\overline{|B|^2}-\mathrm{Re}(A_w)}{D}
\end{align}
with $D=|B|^2-\overline{|B|^2}$. 
Its solution is readily found to be 
\begin{equation}
\label{sol}
\xi_i(k)=\frac{\exp{\left\{-i\langle\Hat{x}\rangle_f k -i\left[\Delta\langle\Hat{x}\rangle\overline{|B|^2}-\mathrm{Re}(A_w)\right]\, H(k)\right\}} 
}{\sqrt{|D(k)|}} .
\end{equation}
with $H(k)$ a primitive of $D(k)^{-1}$.

The parameter $\overline{|B|^2}$ must be found self-consistently by plugging Eq.~\eqref{sol} into 
\begin{equation}
\overline{|B|^2} = \frac{\int dk\, |B\xi_i|^2}{\int dk\, |\xi_i|^2}.
\end{equation}
We have 
\begin{equation}
\overline{|B|^2} = \frac{\int dk\, \frac{|B|^2}{|D|}}{\int dk\, \frac{1}{|D|}}=\overline{|B|^2}+\frac{\int dk \frac{D}{|D|}}{\int dk\, \frac{1}{|D|}}.
\end{equation}
Thus the integration region $[k_-,k_+]$ must include the zeros of $D(k)$, and this implies that $\xi_i(k)$ is not normalizable.

The average $\Delta\langle\Hat{x}\rangle$, however, is well defined even if $\xi_i$ is not normalizable. 
By plugging Eq.~\eqref{sol} into Eq.~\eqref{act} we have an identity. Thus, all the functions given in Eq.~\eqref{sol} 
provide an arbitrary value $\Delta\langle\Hat{x}\rangle$ for the shift. 

In the main text, see Eq.~\eqref{guess}, we have provided a different family of normalizable functions that provide an arbitrary 
$\Delta\langle\Hat{x}\rangle$ as well.

This work was performed as part of the Brazilian Instituto Nacional de Ci\^{e}ncia e
Tecnologia para a Informa\c{c}\~{a}o Qu\^{a}ntica (INCT--IQ) and 
it was supported by Funda\c{c}\~{a}o de Amparo \`{a} Pesquisa do 
Estado de Minas Gerais through Process No. APQ-02804-10.


\begin{thebibliography}{1}
\bibitem{paper}
Y. Susa, Y. Shikano, and A. Hosoya,  Phys. Rev. A \textbf{85}, 052110 (2012).
\end{thebibliography}
\end{document}